\documentclass{emulateapj}
\shorttitle{LEDA 074886}
\shortauthors{Graham et al.}

\begin{document}

\title{LEDA 074886: A remarkable rectangular-looking galaxy}

\author
{Alister W.\ Graham\altaffilmark{1}, 
Lee R.\ Spitler, Duncan A.\ Forbes}
\affil{Centre for Astrophysics and Supercomputing, Swinburne University
of Technology, Hawthorn, Victoria 3122, Australia.}
\author{Thorsten Lisker}
\affil{Astronomisches Rechen-Institut, Zentrum für Astronomie der Universität
  Heidelberg, Mönchhofstraße 12-14, 69120 Heidelberg, Germany.}
\author{Ben Moore}
\affil{Institute for Theoretical Physics, University of Zurich, Zurich, Switzerland.}
\author{Joachim Janz}
\affil{Division of Astronomy, Department of Physics, University of Oulu,
  P.O.\ Box 3000, FI-90014 Oulun Yliopisto, Finland.}
\affil{Astronomisches Rechen-Institut, Zentrum für Astronomie der Universität
  Heidelberg, Mönchhofstraße 12-14, 69120 Heidelberg, Germany.}
\altaffiltext{1}{Corresponding Author: AGraham@astro.swin.edu.au}

\begin{abstract}

We report the discovery of an interesting and rare, rectangular-shaped galaxy.
At a distance of 21 Mpc, the dwarf galaxy LEDA~074886 has an absolute $R$-band magnitude of
$-17.3$ mag.  Adding to this galaxy's intrigue is the presence of
an embedded, edge-on stellar disk 
(of extent $2R_{\rm e,disk} =12\arcsec =1.2$ kpc) for which 
Forbes et al.\ reported $v_{\rm rot}/\sigma \approx 1.4$. 
%
%
We speculate that this galaxy may be the remnant of two (nearly edge-one) merged disk galaxies in
which the initial gas was driven inward and subsequently formed the inner disk, while
the stars at larger radii effectively experienced a dissipationless merger
event resulting in this `emerald cut galaxy' having {\it very} boxy 
isophotes with $a_4/a =-0.05$ to $-0.08$ from 3 to 5 kpc. 
This galaxy suggests that knowledge from simulations of both `wet' and `dry' galaxy
mergers may need to be combined to properly understand the various paths that
galaxy evolution can take, with a particular relevance to blue elliptical
galaxies. 

\end{abstract}


\section{Introduction}

In the universe around us, the overwhelming majority of bright galaxies exist in one
of three main forms (Sandage 1961).  Many, including our own Milky Way galaxy,
are in the shape of a flattened circular disk, typically hosting a spiral
pattern of stars and a central bulge.  A significant number of galaxies are
shaped something like an ellipsoidal football, while the remaining few are
lumpy and irregular in appearance.  Pushing into the dwarf galaxy regime, 
while many elliptical and lenticular dwarf galaxies exist, the 
spiral galaxies tend to give way to more irregular looking galaxies without a
clear symmetry or form.  

We have discovered an exceedingly rare example of a rectangular-shaped galaxy,
which may be unique for its luminosity.  LEDA~074886 ($\alpha$ = 03h 40m
43.2s, $\delta$ = -18d 38m 43s: J2000) resides within the hot gas halo of the
massive, spherical galaxy NGC~1407 (E0), with a projected separation of $\sim$50
kpc from this giant galaxy's core.  LEDA~074886, which we affectionately call 
the ``emerald cut galaxy'' given its striking resemblance to an emerald cut
diamond, is the 25$^{th}$ brightest galaxy within
the NGC~1407 Group of over 250 galaxies (Trentham, Tully, \& Mahdavi 2006).

\section{Data and Analysis}

\begin{figure}
\epsscale{1.0}
\plotone{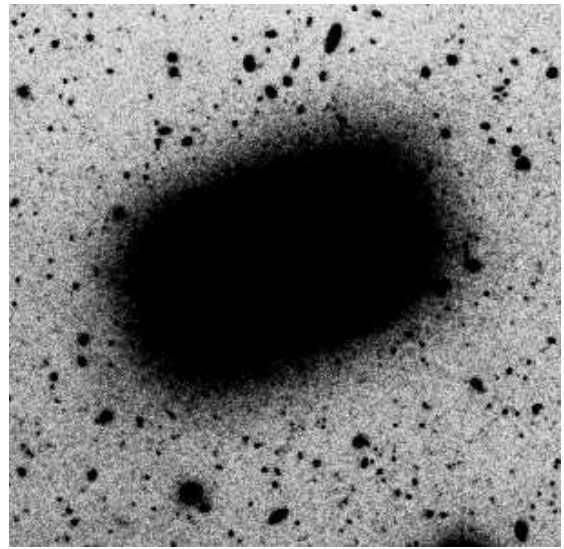}
\caption{
A 120$\times$120 arcsecond ($\sim$12$\times$12 kpc)
$i$-band image of LEDA~074886.
North is up and East is left.  
NGC~1407 is located $\sim$50 kpc to the NW.  
}
\label{Fig1}
\end{figure}

At a distance of 21 Mpc (Spolaor et al.\ 2008), LEDA~074886 has an apparent
(and absolute) R-band brightness of 14.3 (and -17.3) mag (Trentham et al.\
2006), roughly corresponding to a stellar mass of $10^9 M_{\odot}$ (see Forbes et al.\ 2011,
their section~9.1).
The associated spatial scale is such that 1$\arcsec = 101$ pc. 

The unusual shape of LEDA~074886 was discovered by us in deep $i^{\prime}$-band 
exposures acquired with the Suprime-Cam imager 
(Miyazaki et al.\ 2002) 
on the Subaru 8 m
Telescope.  Figure~\ref{Fig1} reveals the remarkable rectangular nature
of the periphery of this galaxy.

\begin{figure}
\includegraphics[angle=270,scale=0.395]{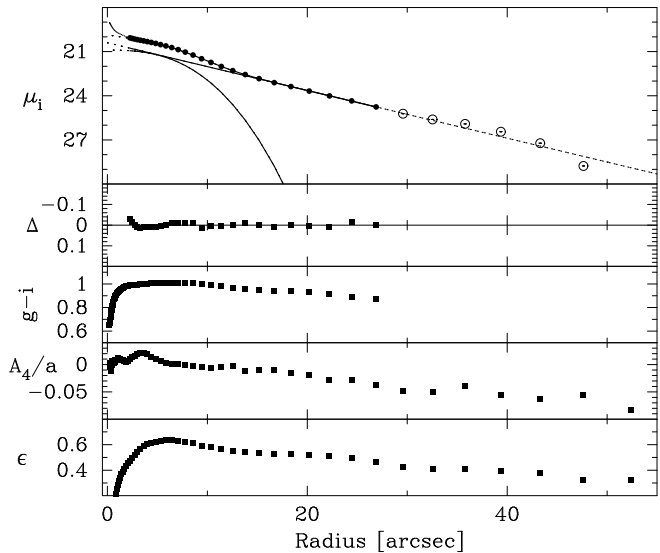}
\label{Fig2}
\caption{
Major-axis, $i^{\prime}$-band light-profile for LEDA~074886 
fit with a S\'ersic model 
($n_{\rm inner}=0.4, R_{\rm e,inner}=5.9\arcsec, \mu_{\rm e,inner}=21.4$ 
mag arcsec$^{-2}$) for the inner embedded disk, plus an exponential
model for the outer host galaxy ($h_{\rm outer}=6.7\arcsec, 
\mu_{\rm 0,outer}=20.4$ mag arcsec$^{-2}$
$\rightarrow R_{\rm e,outer}=11.2\arcsec, \mu_{\rm e,outer}=22.2$
mag arcsec$^{-2}$). 
An additional nuclear component is present, but has been excluded from the
fit over the inner 2$\arcsec$.  
The residuals $\Delta$ about the fit are shown in the second panel.
The $g^{\prime}-i^{\prime}$ colour, $A_4/a$ boxiness parameter, 
and ellipticity are shown in the subsequent panels. 
The scale is such that $10\arcsec = 1.01$ kpc, which is roughly the galaxy's
major-axis half-light radius. 
}
\end{figure}

The {\sc ellipse} task in {\sc iraf} was used to extract this galaxy's major-axis 
surface brightness profile.  In so doing, the centre
was held fixed while the ellipticity and position angle were allowed to
vary.  
Due to the background gradient of halo light from the nearby giant galaxy
NGC~1407, the light from NGC~1407 was first modelled and subtracted from the
Suprime-Cam image.
This was achieved by first subtracting a constant sky-background level determined
from galaxy-free regions of the image, then masking objects around NGC~1407
before running {\sc ellipse} on NGC~1407, and then subtracting this model of
NGC~1407 prior to modelling LEDA~074886 itself.
The resultant light profile for LEDA~074886 is shown in Figure~\ref{Fig2},
where it is apparent that this is not a simple one-component 
galaxy that can be described by a single-component S\'ersic model. 
As revealed by Forbes et al.\ (2011), 
the galaxy possesses an embedded, edge-on stellar disk 
extending to $\sim$12$\arcsec$. 
We have therefore modelled the galaxy's distribution of light with an extended inner and outer
component.  
We avoided the inner three seeing disks (3$\times 0\arcsec.7$) because the 
galaxy is additionally nucleated (see Forbes et al.\ 2011). 
Although we have modeled the light profile from 2--27$\arcsec$, the
extrapolated fit provides a good description to at least 
$\approx$45$\arcsec$ where $\mu_{i^{\prime}}=28$ mag arcsec$^{-2}$
(see Figure~\ref{Fig2}).

Figure~\ref{Fig2} also reveals that the inner disk component is well described
with a S\'ersic $R^{1/n}$ model.  However, as this disk is highly inclined, and given the
tendency for dust to be centrally-concentrated in at least large-scale galaxy disks, it
is perhaps not surprising to find that the S\'ersic index is $\approx$0.4,
i.e.\ less than 1 (the 
value associated with an exponential disk).  Allen et al.\
(2006, their Figure~14) reveal that this is in fact a common result for disks.  The
outer part of the galaxy, or rather that hosting the inner disk, 
is found to be well described with an $n=1$ S\'ersic model.
While the model parameters are given in the caption to Figure~\ref{Fig2},
the galaxy's major-axis half-light radius is around 10$\arcsec$, where
the surface brightness is $i^{\prime} = 21.64, r^{\prime} = 21.97$ 
and $g^{\prime} = 22.64$ mag arcsec$^{-2}$ (obtained from additional
Suprime-Cam images).

The task {\sc ellipse} approximates the isophotes using a Fourier series such that 
\begin{equation}
I(\theta) = I_0 + \Sigma^{\infty}_{n=1}(A_n \cos n\theta + B_n \sin n\thetaθ), 
\end{equation}
where $\theta$ is the azimuthal angle and $I_0$ is the average intensity over
the ellipse.  
Within the output of {\sc ellipse}, the $A_4$ coefficient to the
$\cos 4\theta$ Fourier component is normalised by the semi-major axis length $a$
to give $A_4/a$, which is used for quantifying the isophotal deviation from an
ellipse and is positive (negative) when the isophotes are discy (boxy).  The
intensity-weighted mean value across a galaxy, typically measured from 2
seeing radii to 1.5 half-light radii, is denoted by $a_4/a$ (see Carter 1978,
1987; Bender et al.\ 1988; Hao et al.\ 2006).
For reference, typical (non-dwarf) `boxy' galaxies have $-0.02 < a_4/a < -0.00$, with
extreme values reaching as low as $-0.04$ (e.g.\ Hao et al.\ 2006). 
The inner disk in LEDA~074886 has a peak ellipticity of 0.65 and isophotes with $A_4/a$
peaking at $+$0.022 at 3.5$\arcsec$ (Figure~\ref{Fig2}).  Beyond
where the edge of the inner disk has influence ($> 7.5$--$12\arcsec$), the
$A_4/a$ term becomes increasingly negative (boxy), reaching a value of $-0.05$
by 30$\arcsec$ and $-0.08$ by 52$\arcsec$ at which point the galaxy light
appears to truncate rapidly.  
%



\begin{figure*}
\plottwo{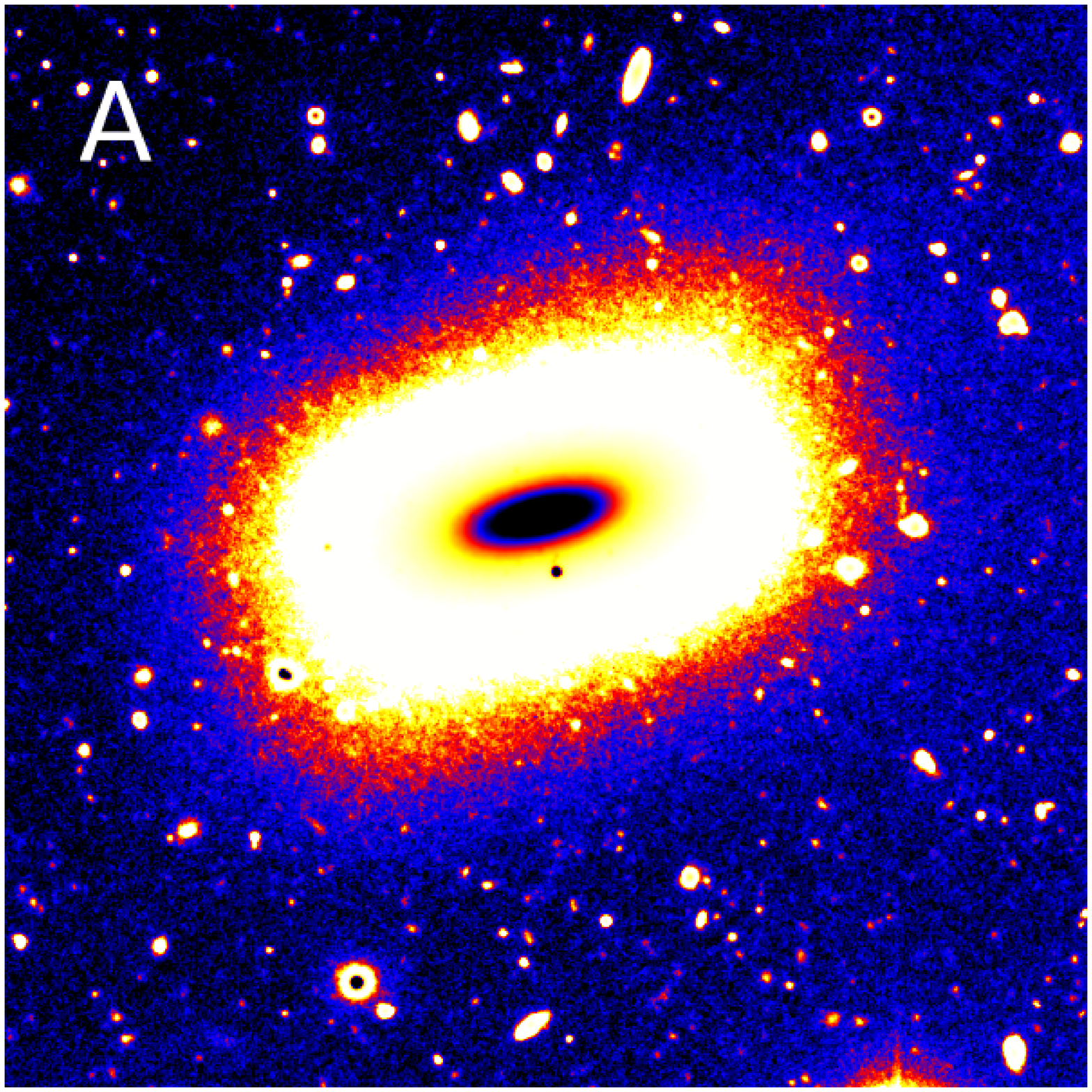}{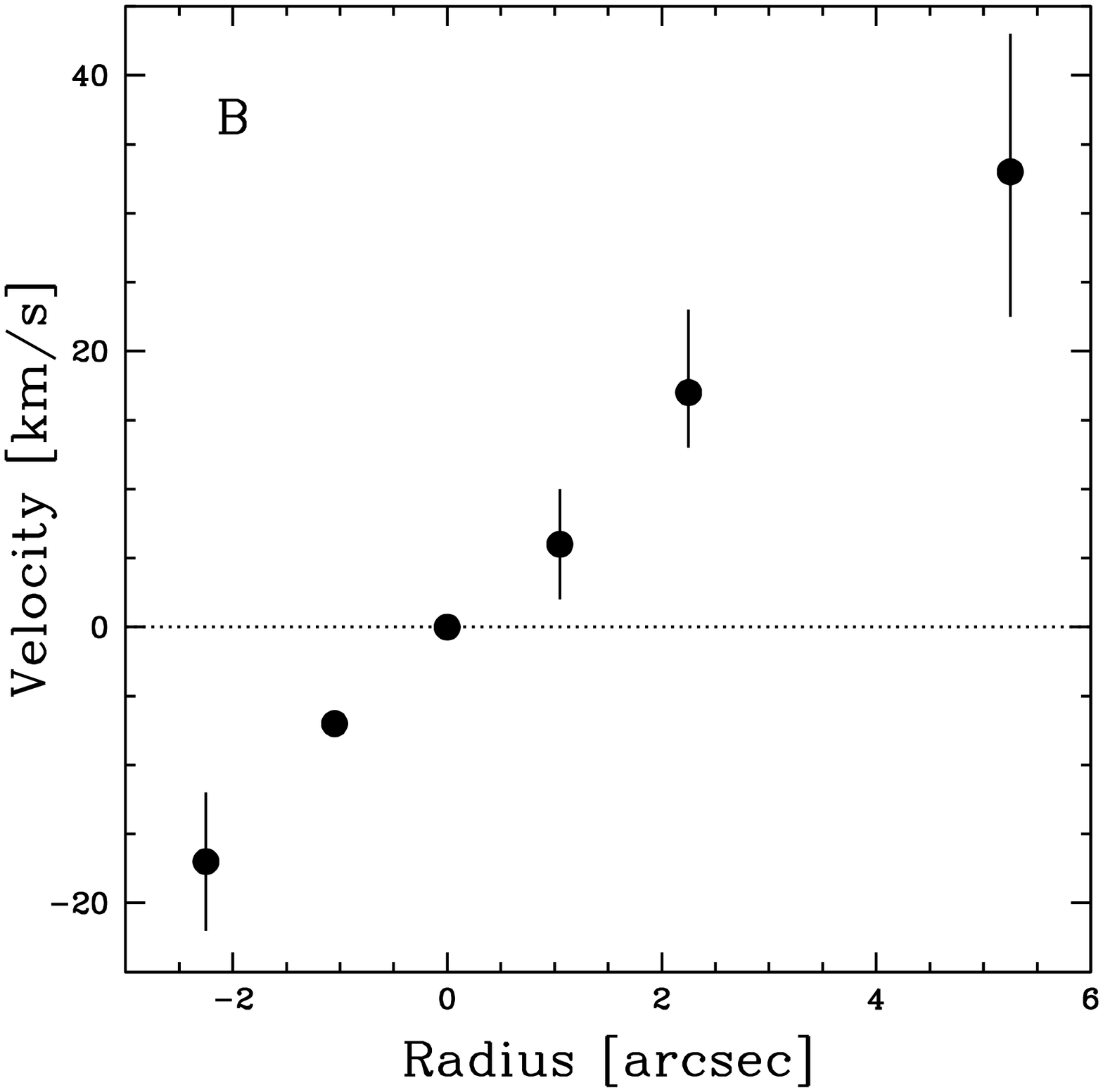}
\caption{
Panel A: 
False-color image of LEDA~074886, taken with Suprime-Cam at the Subaru Telescope. 
The central contrast has been adjusted to reveal the inner disk/bar-like 
component.  For reference, the major-axis of the boxy
outer red annulus spans 3.2--3.8 kpc, while the outer-edge of the outer-most
blue annulus
has a major-axis of 5.2 kpc. 
Panel B: Central rotation curve, taken with ESI at the Keck Telescope, from a
slit placed across the disk of the galaxy; no data is available beyond
$\sim$3 arcseconds (300 pc) in the SE (bottom left) direction.  Profile
adapted from Forbes et al.\ (2011, their Figure~6).
}
\label{Fig3}
\end{figure*}

%

We additionally have Echellette Spectrograph and Imager (ESI: Sheinis et al.\ 2000) 
spectra obtained with the Keck Telescope.  As detailed in Forbes et al.\ 
(2011), this has revealed solid body rotation 
indicating the presence of a central stellar disk (or bar) which reaches a
rotation velocity of 33$\pm$10 km s$^{-1}$ by a radius of $\sim$5 arcseconds
(Figure~\ref{Fig3}b).  This inner most component is evident in 
Figure~\ref{Fig3}a, in which we have adjusted the contrast levels to display
this embedded feature.

\section{Relative Rarity}

In this section we attempt to at least qualify how rare the rectangular
shape of the emerald cut galaxy is.
While common wisdom holds that the Universe tends not to build square-looking
galaxies, there have not been many large quantitative studies on the boxiness of
galaxy isophotes. 
Jiang et al.\ (2011, their Figure~11) did however present single, galaxy
average, indices for 111 galaxies, while Liu et al.\ (2008, their Figure~10)
presented single values for 85 brightest cluster galaxies (BCGs) plus 244
non-BCG early-type galaxies. In general, $a_4/a$ was observed to be confined
to $> -0.02$.  Possibly the largest study to date looking at the distribution
of $a_4/a$ parameters is Hao et al.'s (2006) analysis of 847 luminous
early-type galaxies, in which the bulk have $a_4/a > -0.01$, with just 2.2\%
of the sample having $-0.02 < a_4/a < -0.01$.
However, due to radial changes, i.e.\ gradients, in the boxiness of isophotes,
a single boxy parameter such as $a_4/a$ can be somewhat limited, and will miss
identifying extremely boxy galaxies if only their outer isophotes are boxy, as
is the case with LEDA~074886. 
%

Deep photometry of NGC~3628 reveals an 
edge-on disk galaxy with outer isophotes displaying a rectangular-shaped thick disk, 
see also NGC~4370 (=VCC~758) and NGC~4638 (=VCC~1938). 
From Ferrarese et al.'s sample of 100 early-type Virgo
cluster galaxies, NGC~4638 has the most boxy isophotes, with $a_4/a \sim -0.04$. 
Although whether these three examples of disk galaxies, rather than of
galaxies with embedded disks, can evolve into something 
like LEDA~074886, or vice-versa, is unclear.

\begin{figure}
\epsscale{1.15}
\plotone{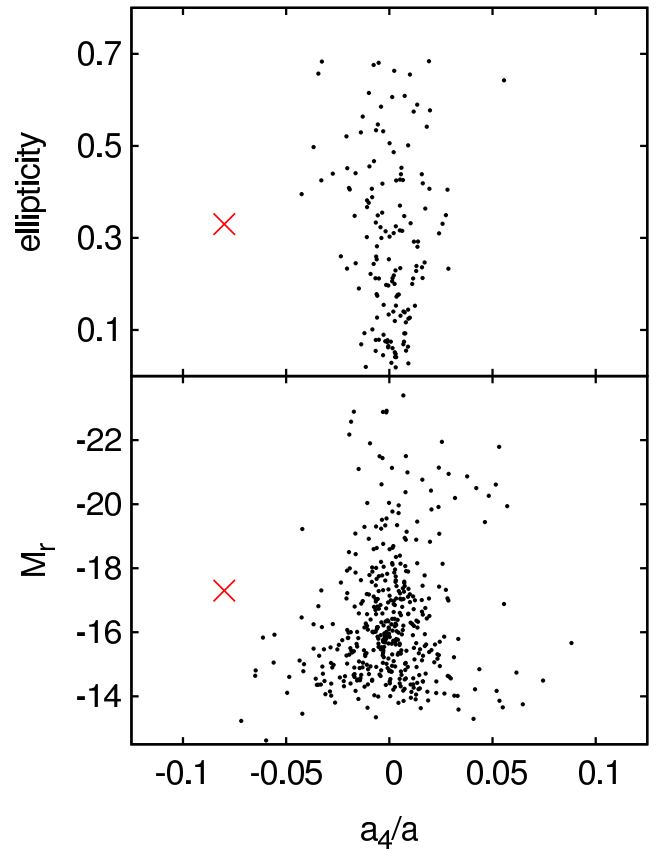}
\caption{The rare nature of LEDA~074886's rectangular-shape 
  can be seen by contrasting its location, denoted by the cross, with that of
  Virgo cluster early-type galaxies in the 
  ellipticty--($a_4/a$) and magnitude--($a_4/a$) diagram.  The 475 Virgo cluster 
  galaxies (lower panel) are described in Janz \& Lisker 2008.  
  For comparison with similarly bright galaxies, the upper panel shows the 
  166 Virgo cluster galaxies with $-19 < M_r < -16$ mag. 
  Note: The $R$-band magnitude for LEDA~074886 ($-17.3$ mag) has been
  used here, along with its ellipticty and $A_4/a$ values at 52$\arcsec$.
}
\label{Fig4}
\end{figure}

The boxiness of our emerald cut galaxy's outer isophotes may be better
appreciated by comparing them with data from 475 early-type Virgo cluster
galaxies taken from Janz \& Lisker (2008).  In Figure~\ref{Fig4} we show the
location of LEDA~074886's outer isophotes in the ellipticty--($a_4/a$) diagram
and the absolute magnitude--($a_4/a$) diagram.  The values for the Virgo
cluster galaxies are roughly the median values from 1.41 to 2.82 $R_{\rm e}$,
rather than the average value within 1.5 $R_{\rm e}$.  
These values, not published before, have come from the {\sc ELLIPSE} fits
generated in Janz \& Lisker (2008). 
The depth of their Virgo 
cluster galaxy images prevents an analysis further out than this.  In future
work we hope to acquire deeper images of the more boxy Virgo galaxies, and
determine how their $A_4/a$ profiles behave with increasing radius.  It should
be noted that the flaring observed at fainter magnitudes in Figure~\ref{Fig4}
is in part due to the increased measurement errors arising from the reduced
signal-to-noise in those galaxy images.
%
%
%
%
In passing it is also noted that the magnitude--($a_4/a$) diagram's branch of `disky'
galaxies at $-20 > M_r > -22$ is populated by lenticular galaxies, while the 
brighter galaxies tend to be boxy (see also, for example, 
Davies et al.\ 1983; Carter 1987; Bender et al.\ 1989; Pasquali, van den Bosch \& Rix 2007).

Due to the limited nature of the above comparison, we additionally asked ourselves what are
the squarish galaxies that we know of. 
Basically, after an extensive literature search, we concluded that 
there are very few square-looking galaxies known in the universe. 
So few in fact that it is possible, and we thought that it may be 
helpful and interesting, to list and contrast them 
with LEDA~074886.  Doing so further highlights the unique nature of this
galaxy. 
\begin{enumerate}
\item{The spiral galaxy {\bf {\sl SDSS} J074018.17+282756.3} has a rather
   squarish-looking spiral-armed interior, but is quite distinct in form to
   LEDA~074886 which has no apparent spiral arms and a boxy exterior.} 
%
\item{A more well known galaxy with a squarish appearance, when looking at the
   expanding shell of young ($<$100 Myr) blue stars, is the Local dwarf
   irregular galaxy {\bf Sextans A} (e.g.\ van Dyk et al.\ 1998).  
   However this galaxy is also very different to
   LEDA~074886, having a much lower mass and density, and no central disk.}
%
\item{The blue compact dwarf galaxy {\bf VCC~1699} (e.g.\ Noeske et al.\ 2003, their
   Figure~12) is another rectangular looking galaxy ($a_4/a = -0.04$),
   although posessing a central comet-like feature with a head-tail
   structure rather than a rotating disk.}
%
\item{The presence of an inner disk in LEDA~074886 suggests that it is not a
  “boxy” elliptical galaxy built solely from a dissipationless major merger (e.g.\
  Naab, Burkert \& Hernquist 1999), as is thought to be the situation with
  the diskless, boxy galaxy {\bf NGC~5322} (Bender, Doebereiner, \& M\"ollenhoff
  1988).  
  The somewhat rectangular-looking 
  lenticular galaxy {\bf NGC~4270} (=VCC 375) has what appear to be outer 
  shells and it may be another example of a dissipationless merger (e.g.\
  Eliche-Moral et al.\ 2006; 
  Gonz\'alez-Garc\'{\i}a et al.\ 2009) with warped, 
  boxy outer isophotes (Gamaleldin 1995), but it too has no inner disk.}
\item{The rectangular / bow-tie shaped galaxy {\bf NGC~4488} (=VCC~1318) also has no
  inner disk; it additionally has two arms coming off diagonally opposite
  `corners' of the main galaxy, suggestive of a gravitational tidal disturbance
  or interaction.}
\item{The (15 degree) misalignment of the inner disk in LEDA~074886 with the
   outer rectangle may rule out the possibility that this is a polar ring
   galaxy like the square-looking {\bf UGC~5119} (Yakovleva \& Karataeva 2004).}
\item{The galaxy with the closest match to LEDA~074886's appearance is 
   not a dwarf galaxy but the somewhat forgotten {\bf IC~3370} (Schweizer \& Ford
   1984).  It is a relatively luminous ($M_B = -20.5$ mag, Galactic extinction corrected)
   $R^{1/4}$ early-type galaxy with a half-light radius $R_{\rm e} = 4$ kpc.}
\end{enumerate}

\section{Discussion}

Efforts to explain the presence of some dwarf spheroidal galaxies within
galaxy clusters, from non-primordial origins, have invoked 'galaxy
harrassment' (Moore et al.\ 1996, 1998, 1999) of disk galaxies.  Through this
process, close encounters can result in pure disk galaxies developing a strong
stellar bar (e.g.\ Noguchi 1988; Gerin, Combes, \& Athanassoula 1990) which
remains intact while the remaining outer disk is eventually stripped away by
multiple high-speed encounters with other galaxies.
Candidates include VCC: 794, 1392, 1400, 1460, 1501, and 1567. 
When seen face-on, these `naked-bars' 
can appear rather rectangular (Mastropietro et al.\ 2005), 
although slightly more elongated than 
LEDA~074886.  A beautiful example of a real `naked bar' can be seen in Penny
et al.\ (2012, in preparation).  However naked bars, when rectangular in appearance, will
have their rotation in the plane-of-the-sky, where as LEDA~074886 displays a
prominant central ``edge-on'' disk rotating perpendicular to the
plane-of-the-sky.

A dwarf galaxy's gas can be removed as it ploughs through the hot halo of
a massive galaxy, preventing any new star formation.  The external tidal
perturbation, strongest during the galaxy's pericenter passages around
the massive galaxy, can also trigger a disk instability which generates a
bar-like component while also kinematically heating many stars out of the disk
plane.  The morphological transformation as the disk, and then bar, puff out
and partially evaporate, both lowers the average stellar density and can result in
the galaxy looking somewhat rectangular (Mayer et al.\ 2007; {\L}okas et al.\ 2010).
However this process does not appear to match the morphology of 
LEDA~074886 with its well-defined embedded disk and outer rectangular morphology. 
%
%
Although cosmological gas inflow, infall events, and secular evolution can
create cold embedded disks, this alone would not explain the boxy nature of
LEDA~074886's outer isophotes.

LEDA~074886 at least {\it resembles} a rotating, vertically-heated disk, i.e.\ a short
cylinder seen side-on, akin to the boxy giant galaxy IC~3370, and if its disk
is precessing then it may explain the misalignment with the outer symmetry. 
Unfortunately LEDA~074886 is too faint for us to acquire dynamical information regarding
the rotational state beyond the inner major-axis.  As such we are unable to
confirm cylindrical rotation, as previously shown for IC~3370, in which the
line-of-sight velocity appears constant along lines running 
parallel to the central rotation axis.

Jarvis (1987a,b) has suggested that IC~3370 may actually be a merger remnant.
Like LEDA~074886, it also has an inner disk, but rotating at a speedy
$\sim$100 km s$^{-1}$.  While IC~3370 displays a 25 degree twist, the position
angle of LEDA~074886 twists from $\sim$100 to $\sim$115 degrees when going
from the inner disk to the outer regions.

The model by May, van Albada, \& Norman (1985) to form peanut/box-shaped
bulges supports a scenario in which a cylindrical torque (possibly from a flyby passage
of NGC~1407) may have transformed a dwarf elliptical galaxy into a
cyclindrically-rotating box-shaped galaxy.  However the presence of an inner
disk in LEDA~074886 suggests that more may be going on than this.

The collision of a disk galaxy with a spheroid-shaped galaxy, or of two
appropriately aligned disk galaxies with major-axes possibly pointing towards the
corners of the box (Binney \& Petrou 1985), may be a better option. 
Such (near) edge-on mergers, coupled with the right viewing angle,  
may explain why LEDA~074886 appears to be so rare.  
Nonetheless, we advocate the exploration of inner disks, and strong negative 
 $A_4/a$ gradients, in other boxy
galaxies such as those plotted in Figure~\ref{Fig4} as it may yield 
new insight into the development of galaxies. 
The simulations by Naab et al.\ (2006) which used inclined mergers of 30 degrees 
generated boxy galaxies.  LEDA~074886 may have formed from disks that merged
in a plane, coupled with gas that dissipated to form the new inner 
disk. This dissipation in the plane would preferentially adiabatically
contract the long axis of the new galaxy.

Building on the work of Jesseit et al.\ (2005) and Naab et al.\ (2006), 
Hoffman et al.\ (2010) have revealed how major, i.e.\ equal mass, mergers of
disk galaxies can, in the presence of 15\%--20\% gas, generate early-type
slowly rotating galaxies with kinematically distinct cores.  They additionally
showed how doubling this gas 
fraction results in galaxies with with sharp embedded disks on the $\sim$1
$R_e$ scale, but which resemble dissipationless mergers, i.e.\ boxy, at large radii not
affected by gaseous star formation (see their Section~3.4). 
Such a hybrid model with star formation in the inner regions of what is effectively a
`dry' merger event at larger radii, may be a plausible explanation for
LEDA~074886, and IC~3370. 

Given LEDA~074886's velocity dispersion of 23 km s$^{-1}$ (Forbes et al.\ 2011), 
one has $v_{\rm rot}/\sigma \sim 1.4$.  
Hoffman et al.\ (2010) report that the inner disks which form in their
gas rich 1:1 disk galaxy merger remnants have 
$v/\sigma$ up to $\sim$1, but they note that 
larger gas fractions (Robertson et al.\ 2006) can produce faster rotating disks.

If the above scenario is correct, then one would expect the inner disk to be
made of younger stars than the surrounding galaxy.  Face-on examples of 
inner disks --- not to be confused with the $\sim$ 10--100 pc nuclear disks
--- which are embedded in a spheroidal-shaped galaxy may be found in NGC~2320
($M_B = -21.7$ mag, Young et al.\ 2009, see their Figure~5) and in IC~3094
(VCC~213, $M_r = -17.5$ mag) --- as seen in the Sloan Digital Sky Survey ({\sl
SDSS}; York et al.\ 2000) repository\footnote{www.sdss.org}.  
%
While the nucleus of LEDA~074886 is already exceedingly blue (see
Figure~\ref{Fig2}), one may additionally expect there to be dust reddening of
the inner disk light.
Assuming that the disk-dominated region of the galaxy is at least as
metal-rich as the outer parts, then the flat colour profile (rather than an
increasingly red color as one heads into the galaxy centre) across the disk
region may be due to a younger mean age for the disk.  This may therefore be
similar to the situation inferred for the disky elliptical galaxy NGC~821
which reveals a luminosity-weighted age of $\sim$4 Gyrs for the disk and
$\sim$12 Gyrs for the outer spherioid stars (Proctor et al.\ 2005).  This
suggests a recent (few Gyrs) dissipative gas event that led to the formation
of new disk stars.
%
%

Martig et al.\ (2009) discussed how red elliptical galaxies containing gas
disks, like NGC 2320 mentioned in the preceding paragraph, may turn into “blue
elliptical” galaxies under certain conditions.  Driver et al.\ (2007) have
shown that these are relatively rare objects, making up just 1.5\% of the
galaxy population brighter than $M_B = -17$ mag, while Lee et al.\ (2008)
report a value of 4\% using {\sl SDSS} DR4 data. 
It is hard not to speculate on the connections between elliptical galaxies
with blue cores (e.g.\ Im et al.\ 2001; Ann 2010) and the presence of inner
disks, and it would be of interest to take a closer look at the blue spheroids
found by Driver et al.\ (2007).  If such galaxies have inner disks which are
dynamically heated and effectively evaporated by the outer spheroid, and if
these disk stars fade to produce a normal red elliptical galaxy, then one may
wonder how many local red elliptical galaxies are descendants of such
high-$z$ blue elliptical galaxies (e.g.\ Shioya \& Yamada 2002) formed from
hybrid wet/dry merger events (e.g.\ Carpineti et al.\ 2012) not involving a galaxy-wide star burst. 

%


\section{acknowledgment}

We are grateful to Nicola Pastorello for his assistance and input.
AWG is supported by Australian Research Council funding (DP110103509 and
FT110100263).
TL is supported within the framework of the Excellence Initiative by the
German Research Foundation (DFG) through the Heidelberg Graduate School of
Fundamental Physics (grant number GSC 129/1).
JJ acknowledges the support of the Gottlieb Daimler and Karl Benz Foundation
and the Academy of Finland.
AWG, DAF and LRS acknowledge financial support from the Access to Major
Research Facilities Programme which is a component of the International
Science Linkages Programme established under the Australian Government's
innovation statement, Backing Australia's Ability.
Based in part on data collected at the Subaru Telescope, which is operated by
the National Astronomical Observatory of Japan, awarded via Gemini Telescope
time swap program GN-2006B-C-18, 
and at the W.M.\ Keck Observatory. 
which is operated by the California Institute of Technology, the University of
California and the National Aeronautics and Space Administration.


\begin{references}
\reference{Allen}Allen, P.D., Driver, S.P., Graham, A.W., et al.\ 2006, MNRAS, 371, 2
\reference{Ann10}Ann, H.B.\ 2010, in Galaxies and their Masks, edited by
Block, David L.; Freeman, Kenneth C.; Puerari, Iv\^anio, Springer Science$+$Business Media, LLC, p.195 
\reference{BDM88}Bender R., Doebereiner S., Moellenhoff C., 1988, A\&AS, 74, 385
\reference{Bend9}Bender R., Surma P., Doebereiner S., Moellenhoff C., Madejsky R., 1989, A\&A, 217, 35
\reference{BaP85}Binney J., Petrou M.\ 1985, MNRAS, 214, 449
\reference{Car12}Carpineti, A., Kaviraj, S., Darg, D., et al.\ 2012, MNRAS, 420, 2139
\reference{Car78}Carter, D.\ 1978, MNRAS, 182, 797
\reference{Car87}Carter, D.\ 1987, ApJ, 312, 514
\reference{Dav83}Davies, R.L., Efstathiou, G., Fall, S.M., Illingworth, G., Schechter, P.L.\ 1983, ApJ, 266, 41 
\reference{Drive}Driver, S.P., Allen, P.D., Liske, J., Graham, A.W.\ 2007, ApJ, 657, L85
\reference{Elich}Eliche-Moral, M.C., Balcells, M., Aguerri, J.A.L.,  Gonz{\'a}lez-Garc{\'{\i}}a, A.C.\ 2006, A\&A, 457, 91 
\reference{Ferr6}Ferrarese, L., C{\^o}t{\'e}, P., Jord{\'a}n, A., et al.\ 2006, ApJS, 164, 334 
\reference{Forbe}Forbes D.A., Spitler L.R., Graham A.W., Foster C., Hau G.K.T., Benson A.\ 2011, MNRAS, 413, 2665
\reference{Gam95}Gamaleldin, A.I.\ 1995, Ap\&SS, 234, 259
\reference{GCA90}Gerin, M., Combes, F., Athanassoula, E.\ 1990, A\&A, 230, 37
\reference{Gon99}Gonz{\'a}lez-Garc{\'{\i}}a, A.C., O{\~n}orbe, J., Dom{\'{\i}}nguez-Tenreiro, R.,  G{\'o}mez-Flechoso, M.{\'A}.\ 2009, A\&A 497, 35 
\reference{Hao06}Hao, C.N., Mao, S., Deng, Z.G., Xia, X.Y., Wu, H.\ 2006, MNRAS, 370, 1339
\reference{Hof10}Hoffman, L., Cox, T.J., Dutta, S., Hernquist, L.\ 2010, ApJ, 723, 818
\reference{Im001}Im, M., et al.\ 2001, Bulletin of the American Astronomical Society, 33, 1180
\reference{JaL08}Janz, J.,  Lisker, T.\ 2008, ApJ, 689, L25
\reference{Ja87a}Jarvis, B.\ 1987a, AJ, 94, 30
\reference{Ja87b}Jarvis, B.\ 1987b, The Messenger, 49, 15
\reference{Jes05}Jesseit, R., Naab, T., Burkert, A.\ 2005, MNRAS, 360, 1185
\reference{Jiang}Jiang, F.-Z., Huang, S., Gu, Q.-S.\ 2011, Research in Astronomy and Astrophysics, 11, 309
\reference{Steli}Kazantzidis, S., {\L}okas, E.L., Callegari, S., Mayer, L., Moustakas, L.A.\ 2011, ApJ, 726, 98
\reference{Klypi}Klypin A., Kravstov A.V., Valenzuela O., Prada F.\ 1999, ApJ, 522, 82
\reference{Lee08}Lee, J.H., Lee, M.G., Park, C., Choi, Y.Y.\ 2008, MNRAS, 389, 1791
\reference{Liu08}Liu, F.S., Xia, X.Y., Mao, S., Wu, H., Deng, Z.G.\ 2008, MNRAS, 385, 23
\reference{Lokas}{\L}okas, E.L., Kazantzidis, S., Majewski, S.R., et al.\ 2010, ApJ, 725, 1516
\reference{Marti}Martig, M., Bournaud, F., Teyssier, R., Dekel, A.\ 2009, ApJ, 707, 250
\reference{Mastr}Mastropietro C., Moore B., Mayer L., Debattista V.P., Piﬀaretti R., Stadel J.\ 2005, MNRAS, 364, 607 
\reference{May85}May, A., van Albada, T.S., Norman, C.A.\ 1985, MNRAS, 214, 131
\reference{Mayer}Mayer L., Kazantzidis S., Mastropietro C., Wadsley J.\ 2007, Nature, 445, 738
\reference{Miy02}Miyazaki, S., Komiyama, Y., Sekiguchi, M., et al.\ 2002, PASJ, 54, 833
\reference{Moo96}Moore, B., Katz, N., Lake, G., Dressler, A., Oemler, A.\ 1996, Nature, 379, 613
\reference{Moo98}Moore, B., Lake, G., Katz, N.\ 1998, ApJ, 495, 139
\reference{Moo99}Moore, B., Lake, G., Quinn, T., Stadel, J.\ 1999, MNRAS, 304, 465
\reference{Naab9}Naab, T., Burkert, A., Hernquist, L.\ 1999, ApJ, 523, L133
\reference{Naab6}Naab, T., Jesseit, R., Burkert, A.\ 2006a, MNRAS, 372, 839
\reference{Nog88}Noguchi, M.\ 1988, A\&A, 203, 259
\reference{PvdBR}Pasquali, A., van den Bosch, F.C., Rix, H.-W.\ 2007, ApJ, 664, 738 
\reference{Proct}Proctor, R.N., Forbes, D.A., Forestell, A., Gebhardt, K.\ 2005, MNRAS, 362, 857
\reference{Rob06}Robertson, B., Bullock, J. S., Cox, T.J., et al.\ 2006, ApJ, 645, 986
\reference{San61}Sandage, A.\ 1961, The Hubble Atlas of Galaxies, Washington: Carnegie Institution, 1961,  
\reference{Sch84}Schweizer F., Ford W.K., Jr.\ 1984, BAAS, 16, 889 
\reference{She00}Sheinis, A.I., Miller, J.S., Bolte, M., Sutin, B.M.\ 2000, Proc.\ SPIE, 4008, 522
\reference{SaY02}Shioya, Y., Yamada, T.\ 2002, in Cosmic Chemical Evolution, edited by K. Nomoto and J. W. Truran. Dordrecht: Kluwer Academic Publishers, IAU Symp.\ 187, 235
\reference{SFHPB}Spolaor M., Forbes D.A., Hau G.K.T., Proctor R.N., Brough S.\ 2008, MNRAS, 385, 667
\reference{Trent}Trentham N., Tully R.B., Mahdavi A.\ 2006, MNRAS, 369, 137
\reference{Dyk98}van Dyk S.D., Puche D., Wong T.\ 1998, AJ, 116, 2341
\reference{YaK04}Yakovleva V.A., Karataeva G.M., Astronomy Letters, 2004, 30(8), 519-526 
\reference{Young}Young, L.M., Bendo, G.J., Lucero M.\ 2009, AJ, 137, 3053
\end{references}
\end{document}